\documentclass[manuscript]{aastex}

\usepackage{longtable}
\usepackage{rotating}
\usepackage{lscape}

\shorttitle{Herschel observed Stripe 82 quasars and their host galaxies.}
\shortauthors{*}

\begin{document}


\title{Herschel observed Stripe 82 quasars and their host galaxies: connections between the AGN activity and the host galaxy star formation}


\author{X. Y. Dong\altaffilmark{1} and Xue-Bing Wu\altaffilmark{1,2}}
\affil{Department of Astronomy, School of Physics, Peking University,
    Beijing,100871, P.R. China}
\affil{Kavli Institute for Astronomy and Astrophysics, Peking University, Beijing,100871, P.R. China}

\email{sunne.xy.dong@pku.edu.cn}

\begin{abstract}

In this work, we present a study of 207 quasars selected from the Sloan Digital Sky Survey quasar catalogs and the Herschel Stripe 82 survey. Quasars within this sample are high luminosity quasars with a mean bolometric luminosity of $10^{46.4}$ erg s$^{-1}$. The redshift range of this sample is within $z<4$, with a mean value of $1.5\pm0.78$. Because we only selected quasars that have been detected in all three Herschel-SPIRE bands, the quasar sample is complete yet highly biased. Based on the multi-wavelength photometric observation data, we conducted a spectral energy distribution (SED) fitting through UV to FIR. Parameters such as active galactic nucleus (AGN) luminosity, FIR luminosity, stellar mass, as well as many other AGN and galaxy properties are deduced from the SED fitting results. The mean star formation rate (SFR) of the sample is 419 $M_{\odot}$ yr$^{-1}$ and the mean gas mass is $\sim 10^{11.3}$ $M_{\odot}$. All these results point to an IR luminous quasar system. Comparing with star formation main sequence (MS) galaxies, at least 80 out of 207 quasars are hosted by starburst galaxies. It supports the statement that luminous AGNs are more likely to be associated with major mergers. The SFR increases with the redshift up to $z=2$. It is correlated with the AGN bolometric luminosity, where  $L_{\rm FIR} \propto L_{\rm Bol}^{0.46\pm0.03}$. The AGN bolometric luminosity is also correlated with the host galaxy mass and gas mass. Yet the correlation between $L_{\rm FIR}$ and $L_{\rm Bol}$ has higher significant level, implies that the link between AGN accretion and the SFR is more primal. The $M_{\rm BH}/M_{\ast}$ ratio of our sample is 0.02, higher than the value 0.005 in the local Universe. It might indicate an evolutionary trend of the $M_{\rm BH} - M_{\ast}$ scaling relation. 

\end{abstract}





\section{Introduction}

The supermassive black hole (SMBH) is a common component residing in many galactic centres. The SMBH mass is known to be correlated with its host galaxy properties, such as the bulge mass and the velocity dispersion \citep[e.g.,][]{Kormendy1995,Magorrian1998,Ferrarese2000,Gebhardt2000,Tremaine2002}. Furthermore, both the cosmic star formation and the black hole accretion activity peak around $z=2$ and decrease towards lower redshifts \citep[e.g.][]{Boyle1998,Madau2014}. It seems that the sub-pc-scaled BH accretion and the kpc-scaled star formation are somehow entwined.  Yet how the SMBH is fuelled and how the star formation is triggered, as well as the interactions between these two processes are still under discussion. 

It is well known that the star formation rate (SFR) and the stellar mass in the star forming galaxy follow a tight correlation. This so-called star forming main sequence (MS) is thought to reflect a large duty cycle of star formation in galaxies and exist over a large range of redshifts from $z=0$ up to $z=7$ \citep{Brinchmann2004,Noeske2007,Elbaz2007,Daddi2007,Daddi2009,Gonzalez2011}. Starburst galaxies are considered as "off-sequence". Differing from MS galaxies, where star formation is caused by internal secular processes, star formation in starburst galaxies is triggered by gas-rich major mergers and has higher star formation efficiency \citep{Daddi2010,Genzel2010}. \cite{Mullaney2012} studied a X-ray selected moderate-luminosity ($L_{x}: 10^{42} - 10^{44} {\rm erg\,s}^{-1}$) AGN sample with $z<3$, and found that $79\pm10$ percent of AGNs reside in massive, normal main-sequence galaxies. It seems that AGN evolution is dominated by non-merger process. Other studies about AGN hosts reach similar conclusions, such as \cite{Kocevski2012}, \cite{Santini2012}, and \cite{Schawinski2011}. Their results provide evidences that  the starburst galaxies only account for 10\% of the cosmic SFR density at $z\sim2$ \citep{Rodighiero2011, Lamoastra2013}. Yet, \cite{Treister2012} found that the mergers are responsible for triggering the most luminous AGNs ($L_{\rm bol}>10^{45}{\rm erg\,s}^{-1}$). These studies imply that host galaxies of moderate luminous AGN and the most luminous AGNs evolve along different paths. Thus, the relations between the AGN luminosity and the SFR are also dependent on the AGN luminosity.

A correlation between the SFR and the AGN luminosity has been found for high luminosity AGNs. For instance, \cite{Schweitzer2006} used PG quasars and found a correlation between the AGN luminosity and polycyclic aromatic hydrocarbon (PAH) luminosity; \cite{Lutz2008} studied 12 $z\sim2$ millimetre-bright type 1 quasars (with optical luminosity $L_{5100}=10^{45}-10^{47} {\rm erg\,s}^{-1}$) and found a correlation between PAH luminosity and $L_{5100}$ ; \cite{Lutz2010} found an increasing of SFR with AGN luminosity at the highest X-ray ($L_{\rm 2-10\,keV}\geq 10^{44} {\rm erg\,s}^{-1}$) luminosities; \cite{Bonfield2011} used optically selected quasars ($10^{45}<L_{\rm bol}<10^{48} {\rm erg\,s^{-1}}$) and found the correlation between the AGN luminosity and SFR. These luminous AGNs are likely to be fuelled by rapid gas infall associated with major mergers of gas-rich galaxies \citep[e.g.][]{Hopkins2006,Hopkins2008,Somerville2008}. 

The SMBH in low and moderate luminosity AGNs are believed to be fuelled by secular processes \citep[e.g.][]{Hopkins2006,Jogee2006,Younger2008}. Many studies focused on these AGNs, yet no strong correlations between AGN luminosity and SFR (or IR luminosity) are found. For example, \cite{Shao2010} found little dependence of far-infrared luminosity on AGN luminosity for $L_{\rm 2-10\,keV}\leq 10^{44}{\rm erg\,s}^{-1}$ AGN at $z>1$; \cite{Mullaney2012} found no relation between $L_{\rm FIR}$ and $L_{\rm x}$ with X-ray selected moderate luminous quasars ($L_x \sim 10^{42} - 10^{44}$ erg s$^{-1}$);  and \cite{Rosario2012} found that $L_{60}$ is independent of $L_{\rm AGN}$ at low accretion luminosities, but shows a strong correlation with $L_{\rm AGN}$ at high accretion luminosities. These studies again indicate that the AGN growth follows two different paths: the low or moderate luminous AGNs evolve through secular processes and  not directly linked to the states of their host galaxies, while high luminosity AGNs evolve through major mergers and might have a direct link between the black hole growth and bulge growth. 

In this paper, using a quasar sample selected from the SDSS Stripe82, we aim to study the AGN activity and the star formation of its host galaxy. In galaxies, most of the radiation from the newly formed stars is absorbed and reemitted at infrared wavelengths. Therefore, the IR luminosity can be used to estimate the SFR. Yet AGN can also heat the dust and bring contamination to the IR luminosity. A popular method to distinguish star formation and AGN activity is to rely on Spectral Energy Distribution (SED) models. Some studies involving SED analysis are \cite{Elvis1994}, \cite{Richards2006}, \cite{Netzer2007}, \cite{Mor2012}, \citeauthor{Leipski2013}\citeyearpar{Leipski2013, Leipski2014}, and \citeauthor{Xu2015a}\citeyearpar{Xu2015a, Xu2015b}. In this work, we use a Python based custom-written SED fitting routine ({\bf available upon inquiry}). The paper is outlined as follows: the second section describes sample selection and supplementary data from multiple surveys; the third section explains each components used in SED fitting and the results; the fourth section discuss physical parameters derived from the SED fitting; the fifth section is discussion, following with a short summary in the last section. 

\section{Sample selection and data collection}\label{sec:data}

The Sloan Digital Sky Survey (SDSS) Stripe 82 covers approximately 270 deg$^2$ area on the celestial equator in the south Galactic cap, spans from 20$^h$ to 4$^h$ in right ascension and -1.25$^\circ$ $\sim$ 1.25$^\circ$ in delineation \citep{AdelmanMcCarthy2007}. It has been repeatedly imaged by the SDSS through 1998 and 2007. Besides the optical imaging survey and the spectroscopic survey, the SDSS Stripe 82 has also been observed extensively by other surveys from X-ray through UV/optical to IR and radio bandpasses. For example, the UKIRT Infrared Deep Sky Survey \citep[UKIDSS;][]{Lawrence2007} and the Two Micron All Sky Survey \citep[2MASS;][]{Skrutskie2006} at the Near-IR, the Wide-Field Infrared Survey \citep[WISE;][]{Wright2010} at the Mid-IR, the Herschel Stripe 82 Survey \citep[HerS;][]{Viero2014} at the Far-IR. Among these surveys, HerS, an imaging survey conducted by SPIRE abroad the Herschel Space Observatory, covers about 79 deg$^2$ area to an average depth of 13.0, 12.9, and 14.8 mJy beam$^{-1}$ (including the confusion limit of 7 mJy) at 250, 350, and 500 $\mu$m, respectively. It is particularly suitable for studying the FIR excess of quasars, which are generally interpreted as contributions from star formation in their host galaxies.

The quasar sample is selected by cross-identifying the HerS catalog with \cite{Shen2011} quasar catalog (DR7Q hereafter), the compiled catalog based on the spectroscopic quasar catalog \citep{Schneider2010} from the SDSS Data Release 7 \citep{Abazajian2009}; and the Sloan Digital Sky Survey quasar catalog from the tenth data release \citep[DR10Q hereafter;][]{Paris2014}. The search radius is 5 arcsec. The HerS catalog is a band-merged catalog with the 250 $\mu$m sources as positional priors and only includes sources with the signal-to-noise ratio greater than 3. In order to achieve a reliable SED fitting at the FIR, we only selected sources detected at all three SPIRE bands from HerS. The resulting sample is highly biased towards luminous FIR sources. Because of this flux limitation, our sample is affected by the Malmquist bias, which we will give more detailed discussion in later sections. There are 226 quasars in the entire sample, where 153 are from DR7Q and 73 from DR10Q. The redshift of the sample is smaller than 4, with a mean redshift of 1.54. The mean flux at 250 $\mu$m is 53.7 mJy, and the average {\it i}-band AB magnitude is 19.4 mag.

\subsection{Optical photometric measurements and spectral parameters} 

The SDSS imaging survey includes five optical bandpasses: {\it u} (3543 \AA), {\it g} (4770 \AA), {\it r} (6231 \AA), {\it i} (7625 \AA), and {\it z} (9134 \AA). The scale plate of the SDSS is 0.4$^{\prime\prime}$/pixel. Some low redshift quasars in the sample show as extended sources on the SDSS images. Therefore, we use \textquotedblleft CMODEL magnitude\textquotedblright\, instead of \textquotedblleft PSF magnitude\textquotedblright. We also remove quasar emission lines (i.e., H$\alpha$, H$\beta$, {Mg\,{\sc ii}}, and {C\,{\sc iv}) by convolving the continuum-substracted spectra with relevant filters from each bandpasses. The SDSS magnitudes are converted to physical fluxes based on \cite{Fukugita1996}. \cite{Shen2011} provides DR7Q with the full-width at half-maximums (FWHMs) of emission lines, such as H$\alpha$, H$\beta$, {Mg\,{\sc ii}}, and {C\,{\sc iv}}. Other useful parameters to our work are the bolometric luminosity, monochromatic luminosities (at 1350\AA, 3000\AA, and 5100\AA, respectively) and virial BH mass. They are used as sanity checks for our results. The SDSS quasar catalog DR10 provides FWHMs of {Mg\,{\sc ii}} and {C\,{\sc iv}} emission lines.

\subsection{FIR/sub-mm photometry}\label{FIR}

FIR data are from Herschel Stripe 82 Survey \citep{Viero2014}, which consists of 79 deg$^2$ of contiguous imaging with the SPIRE instrument \citep{Griffin2010} on the Herschel Space observatory \citep{Pilbratt2010}. The confusion limit is about 7 mJy at all three bands. The point-source catalog of the HerS in the three bands were produced as follow:

\begin{itemize}
\item{Map filtering: to remove large-scale Galactic cirrus. Maps were constructed using the maximum likelihood mapmaker SANEPIC (Signal and Noise Estimation Procedure Including Correlations; \cite{Patanchon2008}), which separates the low-frequency correlated noise from the sky signal, therefore, better preserves the large-scale variations of the sky.}
\item{Source identification: to identify point sources in the filtered 250 $\mu$m image using the IDL software package STARFINDER \citep{Diolaiti2000} with a Gaussian PSF. The FWHMs of the PSF is 18.15, 25.15, and 36.3 arcsec for 250, 350, and 500 $\mu$m, respectively.}
\item{Source photometry: to measure source photometry using a modified De-blended SPIRE Photometry (DESPHOT) algorithm \citep{Roseboom2010,Roseboom2012}.}
\end{itemize}

The band-merged catalog is constructed using 250 $\mu$m sources as positional priors, and only included sources with singal-to-noise ratio greater than 3 at 250 $\mu$m, whose completeness is estimated to be 50\%. 

We adopt flux density measurements directly from the merged catalog of HerS. The average confuse noise is 7 mJy \citep{Viero2014}. As mentioned in \cite{Viero2014}, some sources are in the shallow regions with two scans, its 3 $\sigma$ level corresponding to 31 mJy. While other sources are in deep regions with three scans, whose 3 $\sigma$ level corresponding to 28 mJy. In our sample, 89 sources are in the deep regions, and the remaining sources are in the shallow regions. The mean flux densities at 250 $\mu$m are $56\pm21$ mJy with minimum value of 32 mJy for deep regions, and $54\pm31$ mJy, with minimum value of 30 mJy for shallow regions, respectively.  It is clear that the differences in the mean flux densities and the minimum fluxes due to the number of scans are much smaller than the confusion limit, thus, the uneven coverage of the HerS scan has very limited effects on our sample. Yet as mentioned earlier, because we included only sources with all three bands detections, the sample is highly biased.

\subsection{Near-IR data} 
2MASS is a near-IR imaging survey. It contains three filters: {\it J} (1.25 $\mu$m), {\it H} (1.65 $\mu$m), and {\it K$_{\rm s}$} (2.16 $\mu$m). At the 10-$\sigma$ level, the photometric sensitivity of the point source catalog are 15.8, 15.1, 14.3 mag at $J$, $H$, and $K_s$, respectively. The best image of 2MASS has a FWHM of 2.5$^{\prime\prime}$. We match with 2MASS using a cross-radius of 5$^{\prime\prime}$ and find 90 objects have detections in at least one bandpass. We choose the profile-fit photometry magnitude from the catalog, and convert the Vega-based magnitude to physical flux based on \cite{Cohen2003}. UKIDSS is another near-IR imaging survey. It carries following bandpasses: {\it Y} (1.03 $\mu$m), {\it J} (1.25 $\mu$m), {\it H} (1.63$\mu$m), and {\it K} (2.20$\mu$m).   It is a $K$ band magnitude limited survey with $K$ band depth of 18.4 mag. We match with UKIDSS DR10PLUS via WFCAM science archive using a cross-radius of 5$^{\prime\prime}$. 135 out of 222 quasars find 4-band detections, 18 quasars have at least two band detections, and 8 quasars only have one band detection. We adopt the aperture corrected magnitudes YAPERMAG3, JAPERMAG3, HAPERMAG3, and KAPERMAG3. They are also converted to physical fluxes based on \cite{Hewett2006}.

\subsection{Mid-IF data} 

WISE maps sky using four filters centred at 3.4, 4.6, 12 and 22 $\mu$m, with an angular resolution of 6.1, 6.4, 6.5, and 12.0 arcsec, respectively. 215 out of 222 have matched counterparts in the AllWISE catalog with a match radius of 5$^{\prime\prime}$. We retrieve profile-fitting magnitudes from the catalog and convert them to physical fluxes based on \cite{Wright2010}.

Before SED fitting, the Galactic reddening is corrected based on \citep{Schlegel1998}. The $K$-correction is also applied assuming a power-law SED with index $\alpha_{\nu}=-0.5$. Throughout this paper, we assume cosmological parameters $h$ = 0.7, $\Omega_m$ = 0.3, and $\Omega_{\Lambda}$ = 0.7.

\section{SED fitting}\label{sec:sed}

\subsection{SED fitting components}\label{subsec:fitting}

In general an observed quasar SED can be decomposed to following components: a power-law representing the UV/optical emission from the accretion disk; a torus representing IR emission from the dusty torus; a host galaxy when the contamination from the stellar light is noticeable; and sometimes a FIR excess contributed by star formation. The wavelength range we used for SED fitting is from 0.15 to 500 ${\mu}$m at the rest-frame. The custom-written SED fitting routine uses Python \textquotedblleft lmfit\textquotedblright, a least-squares minimization package with bonds and constraints. The Levenberg-Marqudardt algorithm is used to minimize the $\chi^2$ and provide the standard errors.

\subsubsection{Power-law component} 
The accretion disk emission at the UV/optical regime can be described by a power-law function, $F_{\nu} \propto \nu^{\alpha}$. We extend this component to the NIR as suggested in \cite{Honig2010} where $F_{\nu} \propto \nu^2$, when $\nu \geq 3 \mu m$. The reason to start from 0.15 $\mu$m is to avoid the contamination from the Ly$\alpha$ emission line (see \cite{Richards2006} for the discussion of how emission lines affecting broad-band photometry). The free parameters are the index $\alpha$ and the scale. 

\subsubsection{Torus} 

The Unification Scheme of AGN requires a toroidal region filled with molecular gas and dust to explain the observed broad-line and narrow-line quasars. Dust in the torus is heat up by UV/optical emission from the accretion disk and re-radiates at infrared. This thermal emission from torus dominates the near- to mid-IR emissions of quasars, and peaks around 10-20 $\mu$m. The torus models used in the SED fitting are from clumpy torus models (CAT3D) by \cite{Honig2010}. For each model, torus SEDs are calculated with inclination of 0 to 90 deg with an interval of 15 deg. Because SDSS quasars are mostly Type I AGNs, we only use models with inclination of 0 and 45 deg. The total number of torus models are 480. They can be scaled to match different quasars. Many SED analyses find an extra luminosity bump at 2-4 ${\mu}m$, it is emitted by hot dust at the innermost part of the standard clumpy torus \citep[e.g.,][]{Barvainis1987, Mor2009, Mor2012}. This hot dust emission is modelled by a blackbody component with a temperature of 1300 K, the typical sublimation temperature of hot dust.

\subsubsection{Host galaxy component and internal extinction} 

The observed quasar image is a combination of both AGN and its host galaxy. Depending on the relative intensity of the host to the AGN, a host component could be needed to yield a good SED fit. De-blending a point-source-alike quasar from an extended host is not easy especially at $z>1$. Many quasars also suffer the internal reddening \citep[e.g.,][]{Webster1995,Young2008}, which brings further complications into determining the quasar's SED at the UV and optical bandpasses. 

\cite{Hao2013} developed a \textquotedblleft quasar-galaxy mixing diagram\textquotedblright\, to estimate host galaxy contribution fraction , $f_g$, at 1$\mu$m by using its SED slopes from 1$\mu$m to 3000 \AA\,($\alpha_{\rm opt}$) and from 1 $\mu$m to 3 $\mu$m ($\alpha_{\rm NIR}$) in the rest frame. The reason behind the \textquotedblleft quasar-galaxy mixing diagram\textquotedblright\, is that the Wien tail of the blackbody thermal emission from the hottest dusts starts to outshine the power-law emission from the accretion disk at the optical bandpass. As a result, the spectral energy distributions of a quasar and a galaxy at near 1 $\mu$m are complete different. Quasars, with their SEDs showing a clear dip at 1 $\mu$m, locate on the mixing diagram where $\alpha_{\rm opt}>0$ and $\alpha_{\rm NIR}<0$. Galaxies, with their SEDs peaking around 1-2 $\mu$m, locate on the mixing diagram where $\alpha_{\rm opt}<0$ and $\alpha_{\rm NIR}>0.8$, instead. \cite{Hao2013} also found that objects affected by quasar internal reddening moving along almost perpendicularly to the line that joining the AGN locus and galaxy locus on the mixing diagram. Based on an object's location, the mixing diagram allows us to estimate the host galaxy contribution, $f_g$, as well as its internal reddening ($A_{\rm int}$).  This process is illustrated in Figure \ref{Fig:gal-qso}. Based on the quasar-galaxy mixing diagram, 93 quasars have $f_g>0.1$. We select two galaxy templates from the SWIRE Template Library \citep{Polletta2007}: an Sb galaxy representing the younger stellar population and an elliptical galaxy (E) representing the older stellar population. Galaxy templates are scaled to match $f_g$ throughout the SED fitting. The reddening of the quasar is also given by the quasar-galaxy mixing diagram. In this sample, 134 quasars have noticeable reddening with the average reddening of $0.16\pm0.053$ mag. They are used to correct the internal reddening assuming a SMC reddening law as suggested in \cite{Hao2013}. 	

\subsubsection{Extra dust component} 

The FIR excess of quasar's SED is interpreted as dust heated by star formation from the host. The dust radiation can be approximated as a \textquotedblleft gray-body\textquotedblright radiation, with the emissivity $\nu^{\beta}$, where the emissivity index $\beta$ is related to the physical properties and the environment of dust grains. For instance, \cite{Planck2011} gave the median value of $\beta$ for our Galaxy as 1.8; \cite{Smith2012} found that $\beta$ varies from 1.7 to 2.5 in Galaxy Andromeda using Herschel data. In order to compare FIR temperature with literatures, we fix $\beta$ value to 1.6 as suggested in \cite{Leipski2013}. The free parameters of FIR dust component are temperature and scale.

\subsection{SED results}\label{subset:results}

We apply the SED fitting on our sample. Some quasars are removed from further analysis, we give explanations below.

\subsubsection{Power-law component} 

The mean power-law slope from our result is $0.031\pm0.33$, steeper than the UV/optical spectra index of -0.44 \citep{VandenBerk2001}. The steeper spectra index is because we have corrected the internal reddening of the quasar. A SED fitting without the internal reddening correction yields a mean power-law slope of $-0.38\pm0.32$ instead.

\subsubsection{Host galaxy component} 

As described in the previous section, we estimate the host galaxy fraction, $f_g$, and the internal reddening using the \textquotedblleft quasar-galaxy mixing diagram\textquotedblright\ from \cite{Hao2013}. The mixing diagram assumes that the quasar SEDs are similar to the mean SED of \cite{Elvis1994} and the internal extinction of quasars fellowing the SMC reddening law. \cite{Scott2014} demonstrates that the \cite{Elvis1994} SED template agrees with other templates such as \cite{Richards2006} and \cite{Shang2011}. It also appears to vary little with cosmic evolution or different Eddington ratio \citep{Hao2011}. \cite{Hao2013} demonstrates that there is little difference among generally used extinction curves, such as SMC, LMC, and MW. Here we choose SMC extinction curve because it is more commonly used in quasars \citep{Hopkins2004, Gallerani2010}. The accuracies of $\alpha_{\rm opt}$ and $\alpha_{\rm NIR}$ depend highly on the availability and quality of photometric measurements. The average photometric measurements used for $\alpha_{\rm opt}$ is 8, and 3 for $\alpha_{\rm NIR}$. There are 22 objects with only one photometric measurement within 1 to 3 $\mu$m. \cite{Hao2013} warns against including longer wavelengths ($>3 \mu$m), because they will bring extra uncertainties to NIR. For those 22 objects, we fit the $\alpha_{\rm opt}$ first, then use the extrapolated photometric value at 1 $\mu$m and the measurement from observation to calculate the slope $\alpha_{\rm NIR}$. As expected, these 22 objects have higher mean errors in both $f_g$ and the internal reddening, with $f_g{\rm Err} = \pm 0.12$ and $A_{\rm int}{\rm Err} = \pm 0.06$ mag, comparing to the remaining objects with $f_g{\rm Err} = \pm 0.056$ and $A_{\rm int}{\rm Err} = \pm 0.04$ mag, respectively. Using $f_g$ and $A_{\rm int}$ as priors, we find 93 quasars required a component of host galaxy {\bf(i.e.,~$f_g>0.1$)}. After the SED fitting, 44 quasars can be fitted with an Sb template, the remaining 49 quasars are better suited with an elliptical template. 

\subsubsection{NIR and MIR dust} 

As mentioned before, AGN emission has two luminosity bumps at NIR and MIR, one is around 10-20 $\mu$m from a clumpy torus, the other around 2-4 $\mu$m from hot dust at the innermost region of the standard torus. CAT3D models can describe the clumpy dusty torii, but cannot produce the 2-4 $\mu$m bump. Without an extra hot-dust component, SED fitting will favour the torus models with smaller open-angles, or increase the scale of the torus to match the NIR bump. Therefore, AGN will have a greater contribution towards the FIR, resulting in a colder FIR dust. Adding a hot-dust component helps constraining the torus model, and as a result, indirectly constrains the FIR dust component as well. \cite{Mor2011} shows that a hot-dust component presents in more than 80\% of type I AGNs. We find a slightly higher percentage (89\%) as 202 out 227 quasars need a hot-dust component. 

\subsubsection{FIR dust temperature} 

For high-z quasars, the additional FIR component can be modelled as a gray-body with a temperature of 40-60 K \citep{Leipski2013}. Yet the typical dust temperature at ultra-luminous infrared galaxies (ULIRGs) can be as low as 25-35 K \citep{Hwang2010}. Thus, the FIR dust temperature is allowed to vary from 10 to 60 K. The initial temperature is set to be 44 K, the mean FIR dust temperature of the high-z quasars \citep{Beelen2006, Leipski2013}. To achieve a reliable temperature, the FIR data should be sampled around the peak of the SED at FIR. Yet, with only three Herschel SPIRE bandpasses, it is not always possible depending on the redshift. Using Monte-Carlo simulation to estimate the temperature errors, we find that the mean value of the temperature to the temperature error ratios of our SED fitting is 8.6, while only 8 objects with the ratios smaller than 3. Therefore, we believe that the temperature estimated through SED fitting is robust. Recently, \citet{Ma2015} studied a sample from the optical-selected SDSS quasars and Herschel very wide field surveys. They conducted their FIR SEDs using two methods: a single-temperature modified blackbody spectrum and a set of starburst templates. By comparing our sample with their work, we find 62 common quasars with both temperature to temperature error ratios larger than 3. The effective temperature of our results is about 10 K lower and the FIR luminosity is about 0.2 dex fainter compared to \citet{Ma2015}. It might be due to that we exclude AGN contribution at the FIR. \cite{Beelen2006} detected six high-redshift ($1.8 \leq z \leq 6.4$) optically luminous radio-quiet quasars at 350 $\mu$m using the SHARC II bolometer camera at the Caltech Submillimeter Observatory. They found the mean value of the grey-body temperature was $47\pm3$ K with a dust emissivity index of $\beta=1.6\pm0.1$. The far-infrared luminosities were around 0.6 to 2.2$\times$10$^{13}$$L_{\odot}$. \cite{Wang2008b} observed four $z \geq 5$ SDSS quasars using SHARC-II at 350 $\mu$m. They found the warm dust temperatures were around 39-52 K, and the FIR luminosities of $\sim 10^{13} L_{\odot}$. \cite{Leipski2013} presented 69 QSOs at $z>5$, with a mean cold component temperature of $\sim 50$ K, and a mean value of the FIR emission 10$^{13} L_{\odot}$. The mean temperature of our sample is $33\pm5.2$ K, which is colder than the dust emission in high redshift quasars. It seems that host galaxies of our sample, which have $z<4$, are closer to ULIRGs.
      
\subsubsection{Revised sample} 

Eight quasars, for lack of measurements at 1-10 $\mu$m, are fitted only with a power-law and a gray-body. The mean temperature of the gray-body is  $35\pm8.8$ K. Without constrains at near- and mid-IR, their gray-body temperatures are not trustworthy, thus these eight quasars are excluded from further analysis. Another 12 quasars clearly show some photometric measurement problems. For instance, the $\nu F_{\nu}$ of SDSS $z$ is at least 10 times brighter than UKIDSS $Y$ or 2MASS $J$. These quasars' photometry need more careful examinations, therefore, are also excluded from further analysis at the moment.

In the end, 207 quasars achieve good SED fitting, among them 149 are from DR7Q, the remaining 58 are from DR10Q. They are used to calculate physical parameters in the next section. We compare the revised sample with the original sample, they occupy the same parameter spaces in redshift and {\it i} magnitude. Some SED fitting examples are shown in Figure \ref{Fig:example}. The result parameters along with physical parameters obtained in the next section are provided in an electronic table. The description of this electronic table is given in Table \ref{Tab:sample}.

\section{Physical parameters from SED fitting}\label{subsec:param}

In this section, we discuss individual parameters estimated based on SED fitting.

\subsection{Monochromatic luminosities and bolometric luminosity} 

The monochromatic luminosities can be estimated from SED fitting.  To check whether our fitting result is reasonable at UV/optical, we use DR7Q to compare $L_{1350}$, $L_{3000}$, and $L_{5100}$ with those in \cite{Shen2011}, as shown in Figure \ref{Fig:line-luminosity}. In general, these two results agree very well with each other. We also see a clear tendency towards higher values in our estimates, especially for $L_{1350}$. It is mostly because we considered the internal extinction of quasars. 

Without X-ray data, we use the bolometric luminosity correctors (BCs) provided in \cite{Shen2011} to compute the bolometric luminosity. They are ${\rm BC_{5100} = 9.26}$ ($z<0.7$), ${\rm BC_{3000}=5.15}$ ($0.7 \leq z < 1.9$), and ${\rm BC_{1350}=3.81}$ ($z \geq 1.9$). In Figure \ref{Fig:bol-luminosity}, we compare the result using DR7Q quasars to \cite{Shen2011}. Quasars with non-negligible internal extinctions are in general higher than \cite{Shen2011}. While quasars with negligible internal extinctions agree with \cite{Shen2011}.  Overall, the bolometric luminosities given by two methods match with each other with the mean difference of 0.20 dex. Because quasars from DR10Q do not possess bolometric luminosities from SDSS quasar catalog DR10, we adopt bolometric luminosities from SED fitting to our entire sample for consistence. The mean bolometric luminosity of the sample, LOG($L_{\rm Bol}$(erg s$^{-1}$)), is $46.4\pm0.694$. Its distribution is shown in Figure \ref{Fig:dist}(a).

\subsection{Virial BH mass}

It is common to estimate BH masses based on single-epoch spectra. The assumption is that the broad emission line region (BLR) of an AGN is virialized. Thus, its central BH mass can be computed using the FWHM of the broad emission-line (as a proxy for the virial velocity) and its corresponding continuum luminosity (as a proxy for the BLR radius). DR7Q quasars have virial BH masses from \cite{Shen2011}. But DR10Q quasars only have FWHMs of {Mg\,{\sc ii}}, and {C\,{\sc iv}}. We adopt the continuum luminosities from SED fitting and estimate DR7Q quasars BH masses using \cite{Shen2011} scheme as follows:

\begin{itemize}
\item if $z<0.7$, \begin{equation}{{\rm Log_{10}}(\frac{M_{\rm BH,vir}}{M_{\odot}})=0.910+0.50 {\rm Log_{10}}(\frac{L_{5100}}{\rm 10^{44} erg\,s^{-1}})+2 {\rm Log_{10}}(\frac{\rm FWHM(H\beta)}{\rm km\,s^{-1}})}\label{eq:hbeta}\end{equation} \citep{Vestergaard2006}
\item if $0.7 \leq z<1.9$, \begin{equation}{\mathrm {Log_{10}}(\frac{M_{\rm BH,vir}}{M_{\odot}})=0.740+0.62 {\rm Log_{10}}(\frac{L_{3000}}{\rm 10^{44} erg\,s^{-1}})+2 {\rm Log_{10}}(\frac{\rm FWHM({Mg\,{\scriptstyle II}})}{\rm km\,s^{-1}})}\label{eq:mgii}\end{equation} \citep{Vestergaard2009}
\item if $z \geq 1.9$, \begin{equation}{{\rm Log_{10}}(\frac{M_{\rm BH,vir}}{M_{\odot}})=0.660+0.53 {\rm Log_{10}}(\frac{L_{1350}}{\rm 10^{44} erg\,s^{-1}})+2 {\rm Log_{10}}(\frac{\rm FWHM({C\,{\scriptstyle IV}})}{\rm km\,s^{-1}})}\label{eq:civ}\end{equation} \citep{Shen2010}.
\end{itemize}
For quasars in DR10Q, only Equations \ref{eq:mgii} and \ref{eq:civ} are used. We combine DR7Q and DR10Q and show the BH mass distribution in Figure \ref{Fig:dist}(b). The mean BH mass of our sample is 10$^{8.97\pm0.600}$ $M_{\odot}$.

\subsection{Host galaxy characteristics}

After the removal of the AGN contribution, the FIR luminosity is dominated by the radiation from the young stars heated dust. If we assume that the dust reradiates all of the bolometric luminosity of the starburst, the SFR can be reasonable deduced from the FIR luminosity. In this work, we compute SFR using an equation provided in \cite{Kennicutt1998}: 
\begin{equation}{\frac{\mathrm {SFR}}{1 M_{\odot}\,\mathrm {yr}^{-1}}=\frac{L_{\mathrm {FIR}}}{2.2\times10^{43} \mathrm {erg\,s}^{-1}} }\label{eq:sfr}\end{equation}
where $L_{\rm FIR}$ is FIR (i.e., the gray-body component) luminosity integrated from 8 to 1000 $\mu m$, and the Salpeter initial mass function  (IMF) is assumed. As mentioned in the previous section, 93 quasars have noticeable host galaxy components, they also contribute to star formation. Therefore, we also integrate host component from 8-1000 $\mu$m. We then adjust both SFR and $L_{\rm FIR}$ as the combination of the host and the gray-body components. The mean SFR of our sample is 419 $M_{\odot}\,\rm {yr}^{-1}$, slightly higher than 415 $M_{\odot}\,\rm{yr}^{-1}$ obtained without the adjustment. 

For quasars with a host galaxy component, we estimate their host galaxy mass via colors. The stellar mass-to-light ratio as a function of colors can be expressed as ${\rm Log}_{10}(M/L)=a_{\lambda} + (b_{\lambda} \times {\rm color})$, where $M/L$ ratio is in solar units. In this work, we adopt the coefficients given in Table 7 of \cite{Bell2003}, while the galaxy $g-r$ color and the $K$ band luminosity are derived from the SED fitting. We also modified $a_{\lambda}$ according to a Kennicutt IMF. The mean stellar masses are $10^{10.9} M_{\odot}$ for those quasars with early-type hosts, and $10^{10.5} M_{\odot}$ for those of late-type hosts, respectively. 

In this section, we also give a rough estimate of the host galaxy gas mass. \cite{Scoville2014} states that the long-wavelength Rayleigh-Jeans (RJ) tail of dust emission is nearly always optically thin, therefore, can be used to estimate the ISM mass in galaxies, presumable the dust emissitivity per unit mass and the dust-to-gas ratio can be constrained. Equation 12 of \cite{Scoville2014} gives the flux density measurement at observed frequency $\nu_{\rm obs}$ as follows:
\begin{equation}\label{eq_12}S_{\nu_{\rm obs}} (mJy)= 0.83 \frac{M_{\rm ISM}}{10^{10} M_{\odot}} (1+z)^{4.8}(\frac{\nu_{\rm obs}}{\nu_{850 \mu m}})^{3.8} \times \frac{\Gamma_{RJ}}{\Gamma_{0}} (\frac{Gpc}{d_{L}})^2
\end{equation}
where $M_{\rm ISM}$ is the mass of ISM, $d_L$ is the luminosity distance. $\Gamma_{RJ}$ is the correction factor for departure from the RJ dependence as the observed emission approaches the SED peak in the rest frame; it is given by: 
\begin{equation}\label{eq_11}\Gamma_{RJ}(T_d,\nu_{\rm obs},z) = \frac{h \nu_{\rm obs}(1+z)/kT_d}{e^{h \nu_{\rm obs}(1+z)/kT_d}-1}\end{equation}%
where $T_d$ is the effective dust temperature from the SED fitting. $\Gamma_0=\Gamma_{RJ}(T_d,\nu_{850},0)=0.71$ is a non-negligible RJ departure. Equation \ref{eq_12} can be only used when $\lambda_{\rm rest}\geq250\mu m$. It is to ensure that the wavelength is on the RJ tail and the dust is likely to be optically thin. \cite{Scoville2014} raises cautions about using SPIRE data and the SED fitted temperature to estimate the ISM masses. They point out that, for galaxies at $z=1\sim2$, SPIRE's bandpasses will be near the FIR luminosity peak and not on the RJ tail, thus the dust is not optically thin. In order to use Equation \ref{eq_12}, we use the rest frame flux density at 250 $\mu m$, derived from the SED fitting and revised Equation \ref{eq_12} as follow: 
\begin{equation}S_{\nu_{250\mu m},z=0}(mJy) = 0.83 \frac{M_{\rm ISM}}{10^{10} M_{\odot}} (\frac{\nu_{250\mu m}}{\nu_{850 \mu m}})^{3.6} \times \frac{\Gamma_{RJ}}{\Gamma_{0}} (\frac{Gpc}{d_{L}})^2 
\end{equation}%
where $S_{\nu_{250\mu m}}$ is the rest frame flux density at 250 $\mu m$. The power index of the frequency ratio is changed from 3.8 to 3.6, because we used emissitivity index $\beta=1.6$ through the SED fitting, instead of 1.8 as used in  \cite{Scoville2014}. Equation \ref{eq_11} is revised to  
\begin{equation}\Gamma_{RJ}(T_d,\nu_{250\mu m},z=0) = \frac{h \nu_{250\mu m}/kT_d}{e^{h \nu_{250\mu m}/kT_d}-1}\end{equation}
\cite{Scoville2014} points out that the flux measured near the FIR peak reflects the dust luminosity rather than its mass, therefore, the effective dust temperature derived from the observed SED might not be suitable to use for mass estimate. Take this concern into consideration, the gas masses given in this section are very crude estimates. The mean gas mass is $10^{11\pm0.45} M_{\odot}$.

\section{Discussions}

We select quasars with noticeable hosts {\bf ($f_g>0.1$)} to form a subsample and denoted as HH. The remaining quasars form another subsample and denoted as HW. The summary of the entire sample and the two subsamples are list in Table \ref{Tab:sum}. Because our sample is FIR-selected, it is not surprising that the sample is biased towards gas rich and IR bright systems. The mean IR luminosity (integrated from 8 $\mu$m to 1000 $\mu$m) is $10^{12\pm0.46} L_{\odot}$, indicating that a great many of quasars in our sample are hosted by starbursts. The mean stellar mass is $10^{10.8\pm0.483} M_{\odot}$, similar to the stellar mass of the star-forming galaxies at $z=1\sim2$ \citep{Mullaney2012}. Although the gas mass estimate in the previous section is very crude, we find that the gas mass ($M_{\rm gas}$), the gas depletion timescales ($\tau_{\rm gas} = M_{\rm gas}/{\rm SFR}$), and the gas mass fractions ($M_{\rm gas}/(M_{\rm gas}+M_{\ast})$) are remarkably close to the values of the IR bright sources in \cite{Scoville2014}. We also find that the characteristic parameters list in Table \ref{Tab:sum} show little to no differences among the entire sample and two subsamples. The main difference between the subsamples HH and HW would be the host galaxy stellar mass. Due to the way we conduct the SED fitting, the subsample HH should only include quasars with relatively brighter or more massive host, while quasars in HW might be relatively fainter or less massive.

We draw SFR as a function of redshift in the central plot of Figure \ref{Fig:SFR-z}. The marginal plots that attached to the x-axis and y-axis are the distribution plots of the redshift and SFR respectively. The redshift vs.~SFR plot shows that SFR increases rapidly with increasing redshift. One of the possible explanations is the Malmquist bias due to the 50\% completeness of the HerS catalog. We calculate the mean value of SFR at $z<1$, $1 \leq z<2$, and $z \geq 2$ and the relevant co-moving volume of each redshift bins. The SFR at $1 \leq z <2$ is 3.7 times of the SFR at $z<1$, while the SFR at $z \geq 2$ is 8.5 times of the SFR at $z<1$. At the same time, the co-moving volume only increases about 4.2 times from when $z<1$ to $2 \leq z <3$. It is clear that the increase of SFR along redshift is only partially due to the Malmquist bias. Our results are comparable to \cite{Mullaney2012}, in which the SFR within $1 \leq z<2$ bin is 3.5 times greater than within $z<1$ bin, while the SFR when $z>2$ is 10.3 times greater than when $z<1$. Because the AGN activity and star formation are known to peak at $z\sim2$ \citep[e.g.][]{Behroozi2013,Burgarella2013, Delvecchio2014,Shankar2009}, we then fit the SFR as a function to the redshift within $z<2$ and $z \geq2$ separately. Besides the rapidly increasing SFR when $z<2$, we also see a slight decrease towards higher redshifts at $z>2$.

\cite{Elbaz2011} gives the redshift evolution of the star formation main sequence (MS) where the specific SFR of MS is ${\rm sSFR_{\rm MS}[Gyr^{-1}]=SFR/M_{\ast}}= 26 \times t^{-2.2}_{\rm cosmic}$, and $t_{\rm cosmic}$ is the cosmic time elapsed since the Big Bang in Gyr. It then defines the starburst as ${\rm sSFR_{\rm SB}[Gyr^{-1}]>52 \times t^{-2.2}_{\rm cosmic}}$, where ${\rm sSFR_{\rm SB}}$ is the specific SFR of a starburst. For subsample HH we draw their redshift vs.~sSFR in Figure \ref{Fig:sSFR-z}. The black solid curve is the MS relation given by \cite{Elbaz2011}. There are 80 quasars located above the starburst curve, which is about 38\% of the entire sample. As discussed at the beginning of this section, the main difference between quasars in the subsample HH and HW is their stellar masses. It is natural to deduce that some quasars in HW have similar SFR as those in HH, but have lower stellar masses, thus can also fall in the starburst region. As a result, the proportion of quasars located in starburst region should be larger than 38\% for our sample. \cite{Mullaney2012} study a group of moderate luminosity AGNs. They find 80\% of AGNs are in star-forming systems, while only 10\% are in starburst systems. The $L_x$ of their sample is $10^{42}-10^{44}$ erg s$^{-1}$. The bolometric luminosity of their sample is $10^{43.5}-10^{45.5}$ erg s$^{-1}$, using the bolometric correction value of 22.4 \citep{Mullaney2012}. The mean bolometric luminosity of our sample is $10^{46.4}$ erg s$^{-1}$, about one magnitude higher than \cite{Mullaney2012}. It seems that high luminosity AGNs are more likely located in starburst galaxies. Previous works such as \cite{Coppin2010} also find that a high percentage of moderate-to-high luminosity AGNs (i.e., Log$L_{\rm x} ({\rm erg\,s^{-1}}) > 43.5$) located above the MS. One could argue that our sample is selected based on the Herschel FIR data, which naturally biased towards high SFR system. In later part of this section, we will show that the AGN luminosity is indeed positively correlated to its host galaxy SFR. The lack of higher sSFR at $z>2$ is due to the incompleteness towards the lower stellar mass quasar hosts. The lack of MS at $z<2$ is also obvious, which is the combined effect of the lack of the higher stellar mass galaxies at the lower redshifts and the FIR limited sample.

For subsample HH, the mean value of $M_{\rm BH}$/$M_{\ast}$ ratio is 0.02, higher than the $M_{\rm BH}$/$M_{\rm bulge}$ ratio of 0.005 in the local Universe \citep{Magorrian1998}. It might indicate an evolutionary trend of the $M_{\rm BH} - M_{\ast}$ scaling relation. There have been many studies focus on the evolution of the scaling relation. For instance, \cite{Decarli2010} studied a sample of 96 quasars with redshift up to 3 and found that the $M_{\rm BH}/M_{\rm host,\ast}$ ratio increases by a factor of 7 from $z=0$ to $z=3$. \cite{Merloni2010} used 89 broad line AGN detected in the zCOSMOS survey in the redshift range $1< z < 2.2$ and found that the average black hole to host galaxy mass ratio evolves positively with redshift. \cite{Bennert2011} studied 11 X-ray selected broad-line AGNs in redshift range $1< z < 2$ and a local comparison sample of Seyfert-1 galaxies. They also found a positive relation between the $M_{\rm BH}/M_{\rm host}$ and redshift, where $M_{\rm BH}/M_{\rm host,\ast} \propto (1+z)^{1.15\pm0.15}$. The evolution of the scaling relation is expressed as follows:
\begin{equation}\label{eq_scaling}{\rm log} M_{\rm BH}-8 = \alpha ({\rm log} M_{\rm host,\ast}-10) + \beta {\rm log}(1+z) + \gamma + \sigma\end{equation}
where $\alpha = 1.12$, the slope of the relations at $z=0$, is assumed not to evolve; $\beta = 1.15\pm0.15$ is used to describe the evolution of the relation; $\gamma = -0.68$ is the intercept of the relation at $z=0$; $\sigma = 0.16\pm0.06$, the intrinsic scatter, is also assumed not to evolve. Using the quasars within subsample HH, we compare the $M_{\rm BH}$ vs. $M_{\ast}$ relation to AGNs in the local universe in Figure \ref{Fig:mhost-mbh}. The black solid line is Equation \ref{eq_scaling} at $z=0$ with $3\,\sigma$ boundary. 42 quasars are located above the $3\,\sigma$ boundary, their BH masses are larger than those of local AGNs with the same host masses. Based on the SED fitting results, subsample HW includes 114 quasars with non-detectable hosts. It is reasonable to deduce that some quasars in subsample HW have relatively fainter hosts than those in HH, therefore, their $M_{\rm BH}$/$M_{\ast}$ ratios might also be higher than those derived from Equation \ref{eq_scaling}. 
Contrary to our results, \cite{Mullaney2012} find that the $M_{\rm BH}/M_{\ast}$ of moderate luminous AGNs is $(1-2)\times10^{-3}$, comparable to the local galaxies. It seems that the coupling between the BH accretion and the star formation is somewhat related to the AGN luminosity, the regulation of the SMBH and its host galaxy bulge has yet to be established in the host galaxy of our highly luminous quasar sample. 

As discussed in the introduction, a strong correlation between SFR and AGN activity can be found in luminous AGN systems but not in low-to-moderate luminosity AGNs. Considering the tight relation between the SFR and $M_{\ast}$ in MS galaxies, the host galaxy stellar mass could play a role in the star formation and AGN activity relation even in galaxies beyond the MS. The galaxy gas providing fuelling for both AGN accretion and star formation might also affect the evolution of these two processes. The Pearson correlation coefficient of $L_{\rm FIR}$ vs.~$L_{\rm Bol}$, $M_{\ast}$ vs.~$L_{\rm Bol}$, and $M_{\rm gas}$ vs.~$L_{\rm Bol}$ are 0.70, 0.53, and 0.50, respectively, with $p$-value $<0.05$. It seems that the SFR is linked more closely with the AGN activity than the stellar mass or gas mass. We plot the FIR luminosity vs.~AGN bolometric luminosity in Figure \ref{Fig:bol-fir}. Their correlation can be expressed as $L_{\rm FIR} \propto L_{\rm Bol}^{0.46\pm0.03}$. The Pearson correlation coefficients of individual subsample HH and HW are 0.72 and 0.68, respectively. It seems that the two subsamples follow the same relation and with the same significant level. Considering the two subsamples share similar SFR and $L_{\rm Bol}$, yet different $M_{\rm host}$, it also implies a closer link between the SFR and the AGN activity than the host galaxy stellar mass. Our results might support the finding in \cite{Delvecchio2015}, that the SFR is the original driver of the correlation between the star formation and the AGN activity.
 
\section{Summary}

Based on the Sloan Digital Sky Survey quasar catalogs, we selected a sample of galaxies that have also been observed by Herschel Stripe 82 survey. One of the main selection criteria was that the sources have been detected by Herschel-SPIRE in all three bands. As a result, the sample is complete yet highly biased towards mid-to-far infrared luminous objects. We conducted a full SED fitting from UV/optical to FIR. Physical parameters were calculated from SED fitting results. The main results are as follow:
\begin{itemize}
\item{The mean SFR is $419 M_{\odot}$ yr$^{-1}$, the mean FIR luminosity is $10^{12.4} L_{\odot}$, similar to the local massive star forming galaxies \citep{Kennicutt1998}. The locations of the quasar hosts on the MS diagram show that at least 26\% quasars are hosted in starbursts.}
\item{The SFR increases with the increasing of redshift and peaks around $z=2$.}
\item{A positive relation between the AGN bolometric luminosity and the FIR luminosity, $L_{\rm FIR} \propto L_{\rm Bol}^{0.46\pm0.03}$, is found . Studies such as \cite{Lutz2010} and \cite{Bonfield2011} also found correlations between the SFR and the AGN luminosity. It indicates that the SFR and the AGN activity are correlated in high luminosity AGNs.}
\item{The AGN bolometric luminosity is also correlated with the host stellar mass and the gas mass, yet with a less significant level than the relation between $L_{\rm FIR}$ and $L_{\rm Bol}$. It agrees with the result in \cite{Delvecchio2015}, that the SFR is an original driver of the connection between the AGN activity and star formation.}
\item{Comparing with the local Universe, the higher $M_{\rm BH}/M_{\ast}$ ratio indicates an evolutionary trend of the $M_{\rm BH} - M_{\ast}$ scaling relation. It seems that in high luminosity AGN systems, the $M_{\rm BH} - M_{\ast}$ scaling relation has yet to be established.}
\end{itemize}

\section*{Acknowledgements}

XYD and XBW thank Luis Ho for his helpful suggestions, and Linhua Jiang for providing Stripe 82 stacked image. 

We thank the supports by the NSFC grants No.~11373008 and No.~11533001, the Strategic Priority Research Program \textquotedblleft The Emergence of Cosmological Structures\textquotedblright of the Chinese Academy of Sciences, Grant No.~XDB09000000, and the National Key Basic Research Program of China, 2014CB845700.

Funding for the SDSS and SDSS-II has been provided by the Alfred P. Sloan Foundation, the Participating Institutions, the National Science Foundation, the U.~S. Department of Energy, the National Aeronautics and Space Administration, the Japanese Monbukagakusho, the Max Planck Society, and the Higher Education Funding Council for England. The SDSS Web site is \url{http://www.sdss.org/.}
The SDSS is managed by the Astrophysical Research Consortium for the Participating Institutions. The Participating Institutions are the American Museum of Natural History, Astrophysical Institute Potsdam, University of Basel, University of Cambridge, Case Western Reserve University, University of Chicago, Drexel University, Fermilab, the Institute for Advanced Study, the Japan Participation Group, Johns Hopkins University, the Joint Institute for Nuclear Astrophysics, the Kavli Institute for Particle Astrophysics and Cosmology, the Korean Scientist Group, the Chinese Academy of Sciences (LAMOST), Los Alamos National Laboratory, the Max-Planck-Institute for Astronomy (MPIA), the Max-Planck-Institute for Astrophysics (MPA), New Mexico State University, Ohio State University, University of Pittsburgh, University of Portsmouth, Princeton University, the United States Naval Observatory, and the University of Washington.
Funding for SDSS-III has been provided by the Alfred P. Sloan Foundation, the Participating Institutions, the National Science Foundation, and the U.S. Department of Energy Office of Science. The SDSS-III web site is \url{http://www.sdss3.org/.}
SDSS-III is managed by the Astrophysical Research Consortium for the Participating Institutions of the SDSS-III Collaboration including the University of Arizona, the Brazilian Participation Group, Brookhaven National Laboratory, Carnegie Mellon University, University of Florida, the French Participation Group, the German Participation Group, Harvard University, the Instituto de Astrofisica de Canarias, the Michigan State/Notre Dame/JINA Participation Group, Johns Hopkins University, Lawrence Berkeley National Laboratory, Max Planck Institute for Astrophysics, Max Planck Institute for Extraterrestrial Physics, New Mexico State University, New York University, Ohio State University, Pennsylvania State University, University of Portsmouth, Princeton University, the Spanish Participation Group, University of Tokyo, University of Utah, Vanderbilt University, University of Virginia, University of Washington, and Yale University.

This publication makes use of data products from the Wide-field Infrared Survey Explorer, which is a joint project of the University of California, Los Angeles, and the Jet Propulsion Laboratory/California Institute of Technology, funded by the National Aeronautics and Space Administration.



\clearpage

\begin{figure}
\plotone{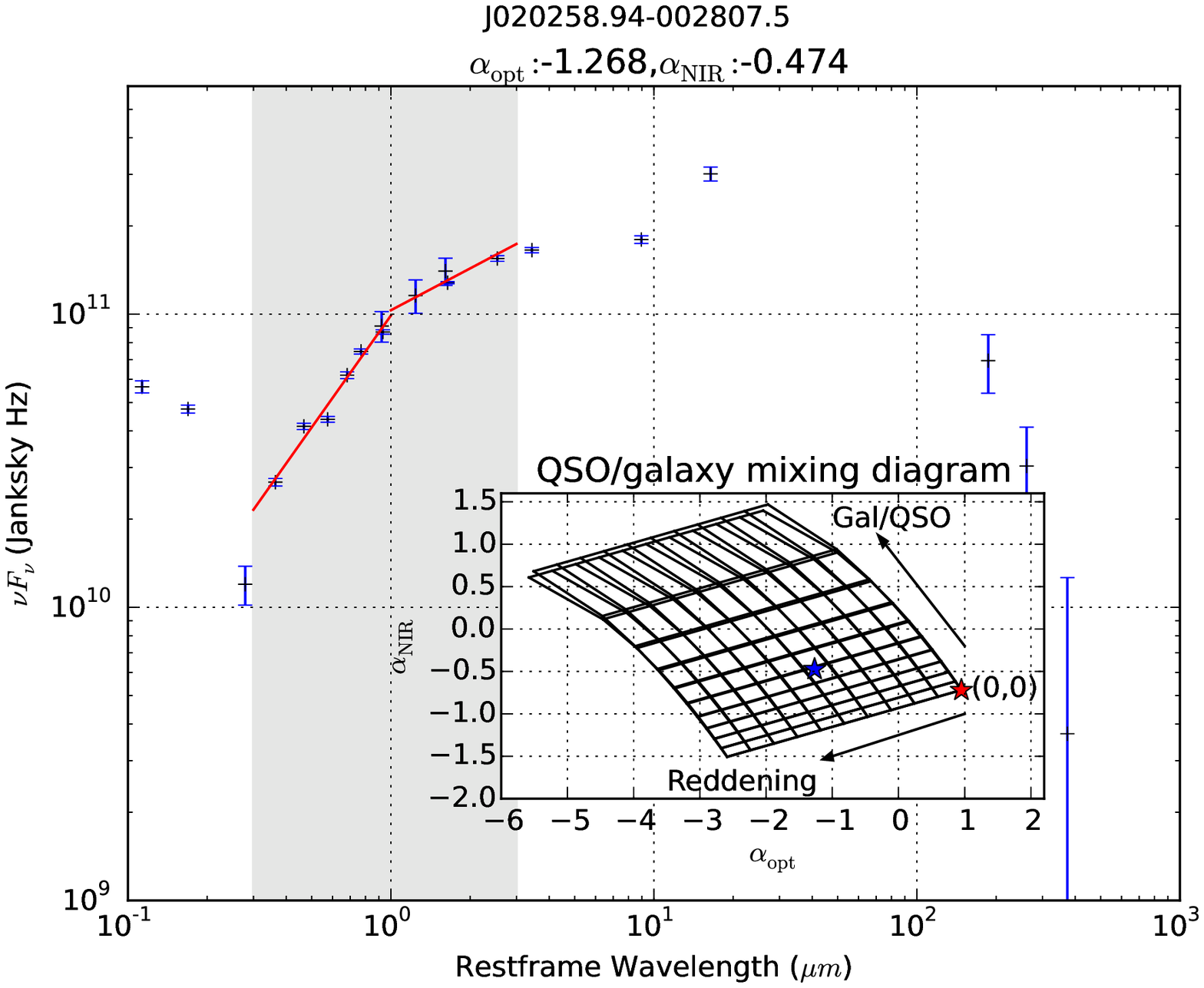}
\caption{An example for estimating the host galaxy contribution fraction and the internal reddening using $\alpha_{\rm opt}$ and $\alpha_{\rm NIR}$. The quasar broadband photometric measurements are represented by black crosses with blue error bars. $\alpha_{\rm opt}$ is the slope from 0.3 to 1 $\mu$m. $\alpha_{\rm NIR}$ is the slope from 1 to 3 $\mu$m. The inset is the quasar-galaxy mixing diagram. The grids are generated for an Sb and an elliptical galaxies assuming a SMC reddening law. The far right corner of the grid, marked with a red asterism, represents the quasar mean SED with both $f_g$ and reddening equal to zero. The $f_g$ increases towards the upper left corner with an interval of 0.1. The reddening increases towards the lower left corner with an interval of 0.1 mag. The blue asterism indicates the location of the quasar on the quasar-galaxy mixing diagram, which yields a $f_g$ of 0.5 and reddening of 0.5 mag. \label{Fig:gal-qso}}
\end{figure}

\begin{figure}
\plotone{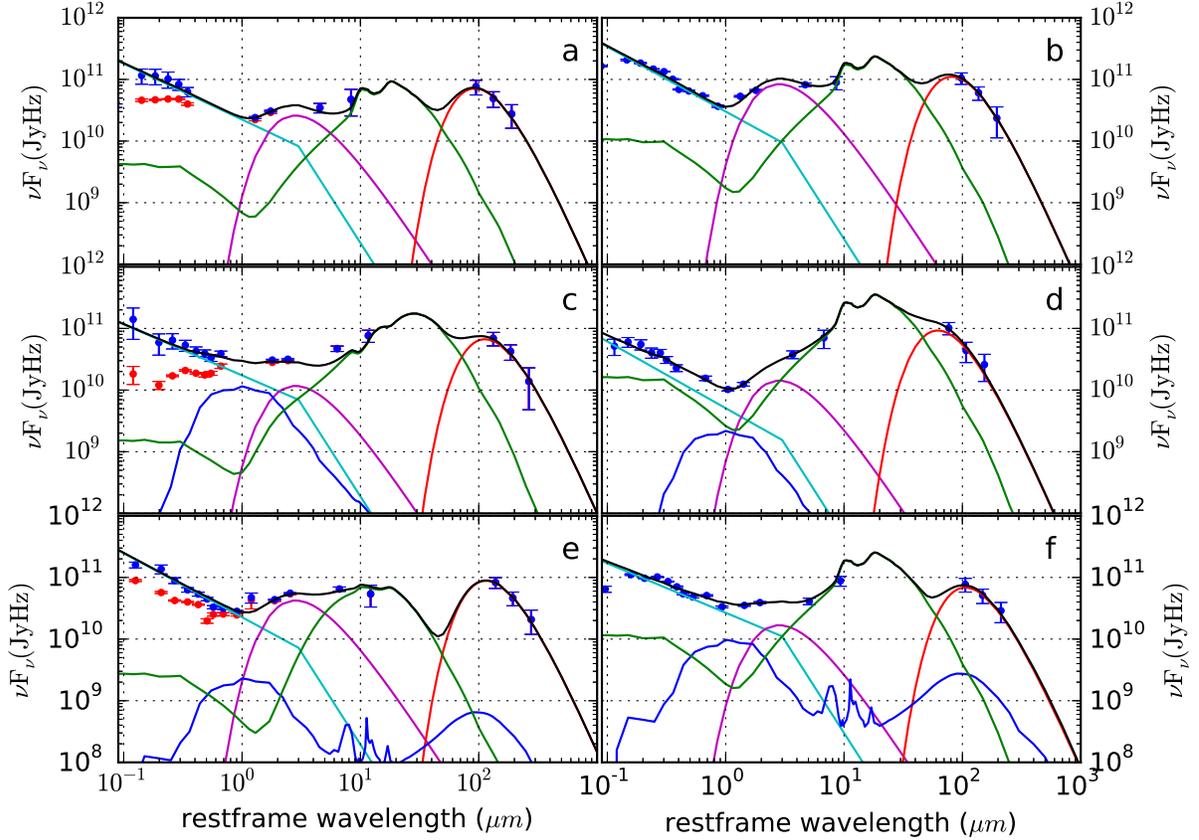}
\caption{A few examples for the SED fitting. The left column are quasars with non-negligible internal reddening. The right column are quasars with negligible internal extinction. The top two plots are quasars that do not need host components. The middle two plots are quasars with relatively older stellar populations. The bottom two plots are quasars with relatively younger stellar populations. In each graph, the red dots with error bars are broadband photometric measurements. The blue dots with error bars are after the internal reddening correction. The cyan curve is the power-law component. The magenta curve is the hot dust component. The green curve is the dusty torus component. The red curve is the cold dust component. The blue curve is the host galaxy component. The black curve is the combination of all components. The detailed information for these quasars are given in Table \ref{Tab:example}. \label{Fig:example}}
\end{figure}

\begin{figure}
\plotone{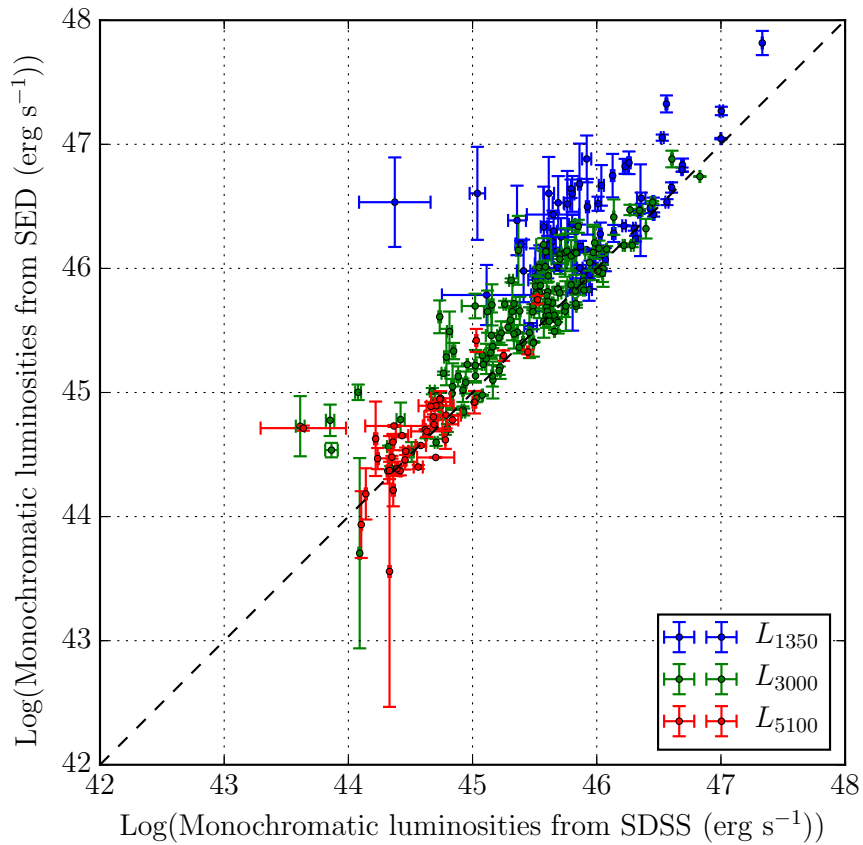}
\caption{The comparison of the monochromatic luminosities estimated via SED fitting and from \cite{Shen2011}. The dashed diagonal line denotes that the two values are equal. The two estimations agree mostly with each other with our estimates slightly higher than \cite{Shen2011}, especially for $L_{1350}$, because we considered the internal extinction of quasars.
\label{Fig:line-luminosity}}
\end{figure}

\begin{figure}
\plotone{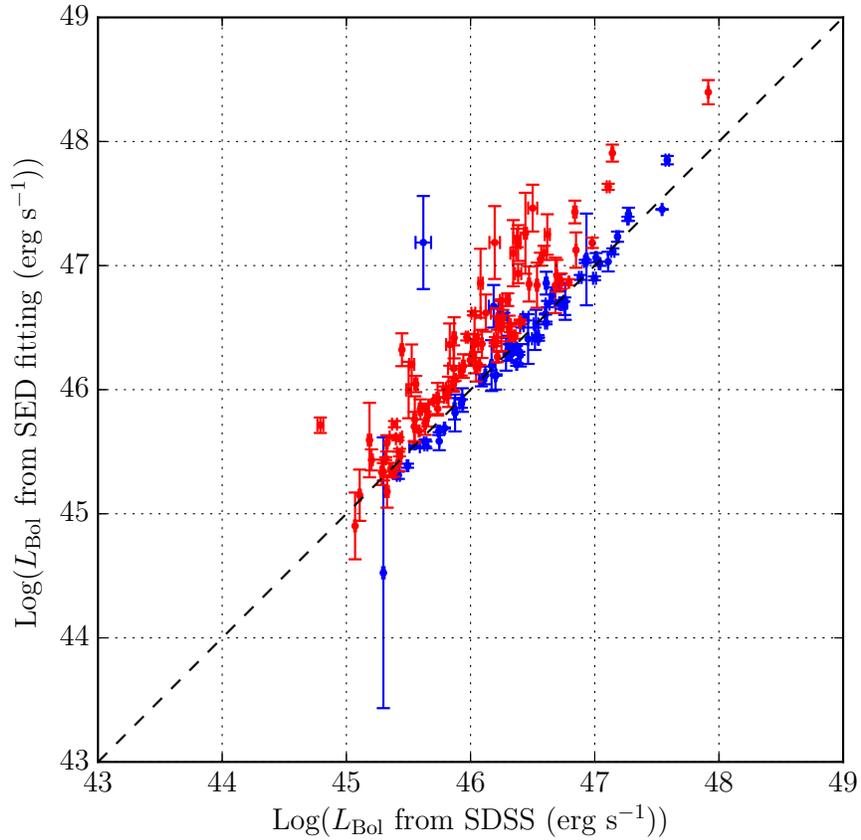}
\caption{The comparison of the bolometric luminosities estimated via SED fitting and from \cite{Shen2011}. The dashed diagonal line denotes that the two values are equal. The data points in red represent quasars with non-negligible internal extinction, while data points in blue are quasars with negligible internal extinction. It is clear that the slightly higher values of some of our estimates are because we corrected the internal extinction. 
\label{Fig:bol-luminosity}}
\end{figure}

\begin{figure}
\plotone{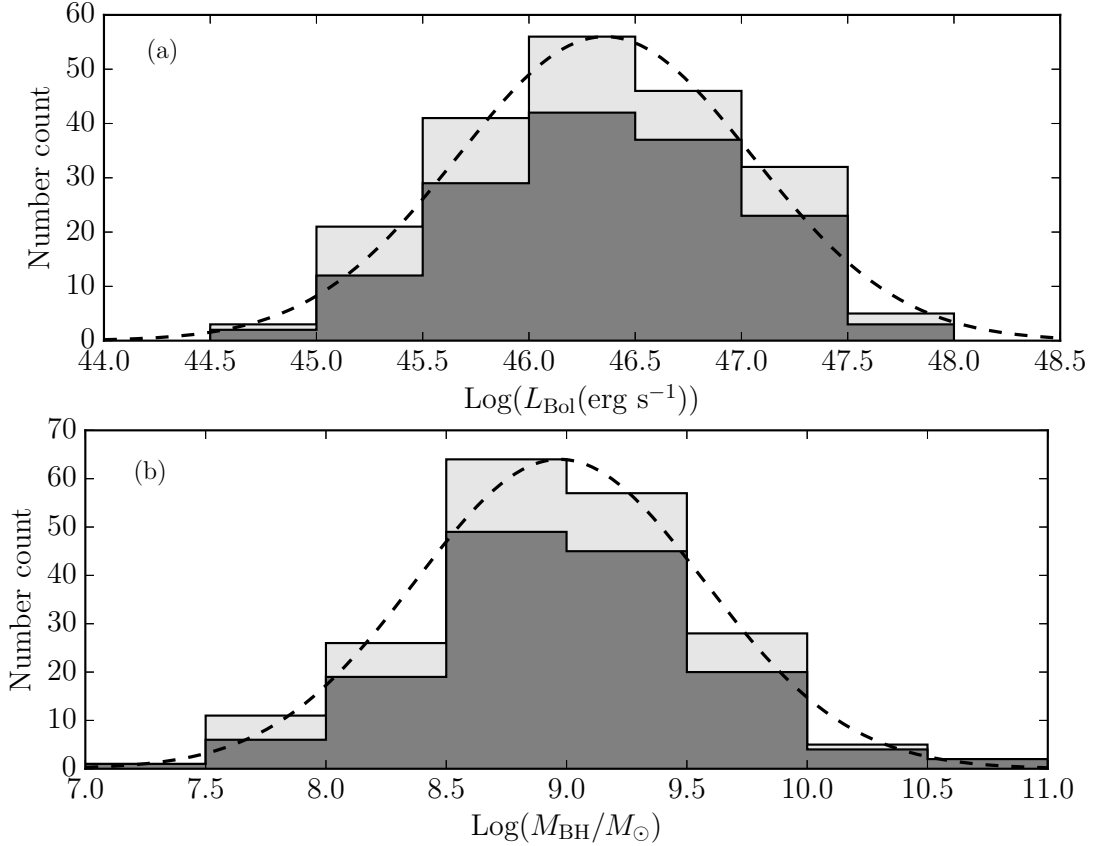}
\caption{(a).~The distribution of the bolometric luminosity estimated via SED fitting. DR7Q is represented  by dark-grey shade, DR10Q by light-grey shade. The dashed curve represents the entire sample. (b).~The distribution of the SMBH mass computed with line luminosities from SED fitting. DR7Q is represented  by dark-grey shade, DR10Q by light-grey shade. The dashed curve represents the entire sample.
\label{Fig:dist}}
\end{figure}

\begin{figure}
\plotone{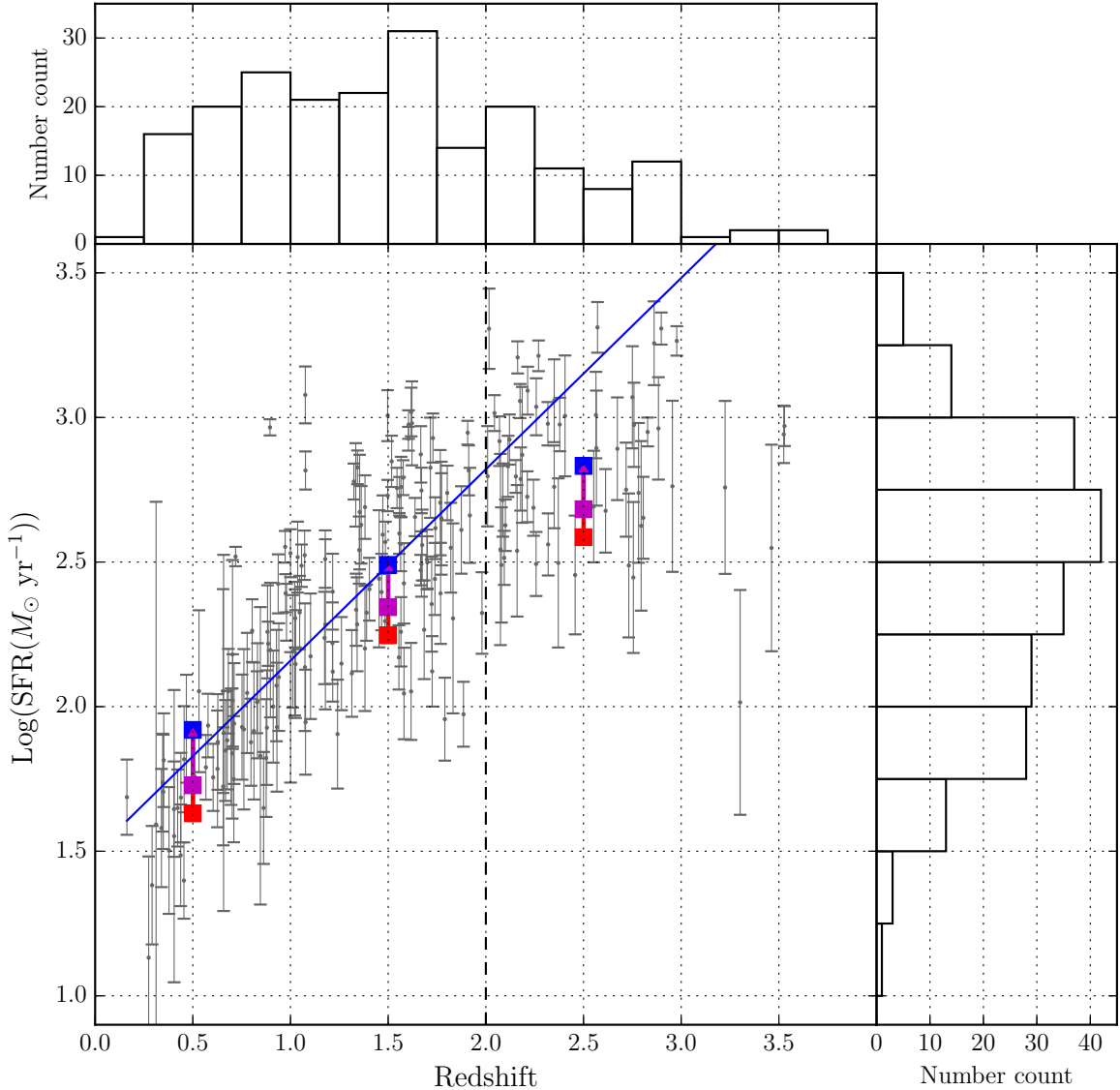}
\caption{\footnotesize The redshift and SFR. The central plot shown redshift vs.~SFR relation. The vertical black dotted-dashed line marks $z=2$. We fit quasars as a function of redshift when $z<2$. The SFR of $z<2$ quasars (blue line) increases rapidly towards $z=2$, while the SFR of $z \geq 2$ quasars decreases slightly. The blue squares are the mean SFRs at $z<1$, $1 \leq z<2$,  and $z \geq 2$, respectively. We also estimated the SFR corresponding to the $L_{\rm FIR}$ limit within each redshift bin. The red squares represent areas with 3 scans, which corresponding to 28 mJy with 3$\sigma$ detection at 250 $\mu$m, the magenta squares represent areas with 2 scans, which corresponding to 31 mJy with 3$\sigma$ detection at 250 $\mu$m. The marginal plot attached to the x-axis is the distribution plot of redshift. The marginal plot attached to the y-axis is the distribution plot of SFR. A Malmquist bias can be seen in each plot.\label{Fig:SFR-z}}
\end{figure}

\begin{figure}
\plotone{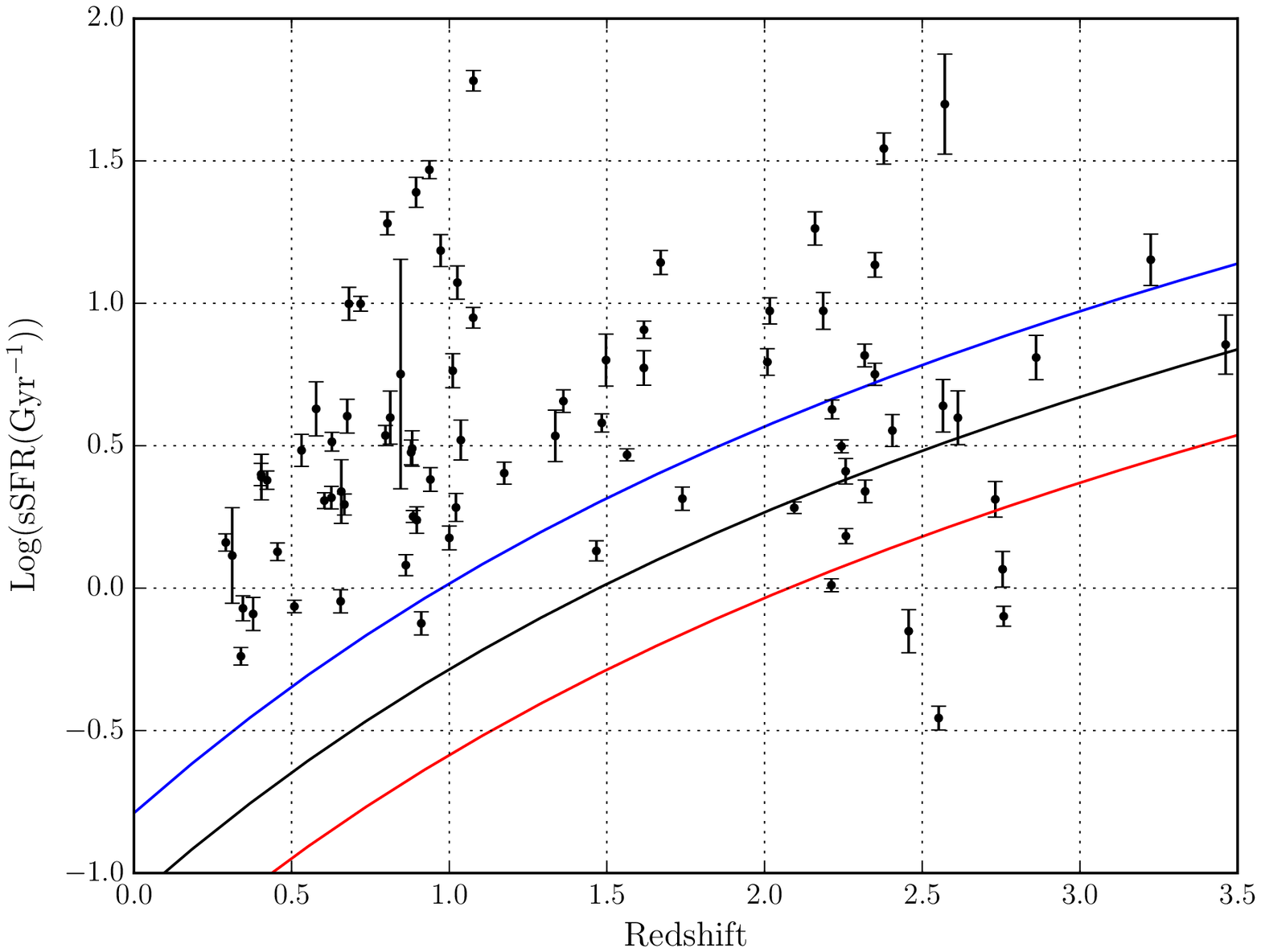}
\caption{The redshift vs.~sSFR (Gyr$^{-1}$) of subsample HH. The black solid curve indicates the star formation main sequence (MS), where ${\rm sSFR}_{\rm MS}({\rm Gyr}^{-1}) = 26\times t^{-2.2}_{\rm cosmic}$. The starburst follows the blue solid curve, where ${\rm sSFR_{\rm MS}(Gyr^{-1}) }= 2\times26\times t^{-2.2}_{\rm cosmic}$ \citep{Elbaz2011}. The red solid curve is a factor of 2 below the main sequence curve. There are 80 objects located above the starburst curve, which is about 38\% of the entire sample. \label{Fig:sSFR-z}} 
\end{figure}

\begin{figure}
\plotone{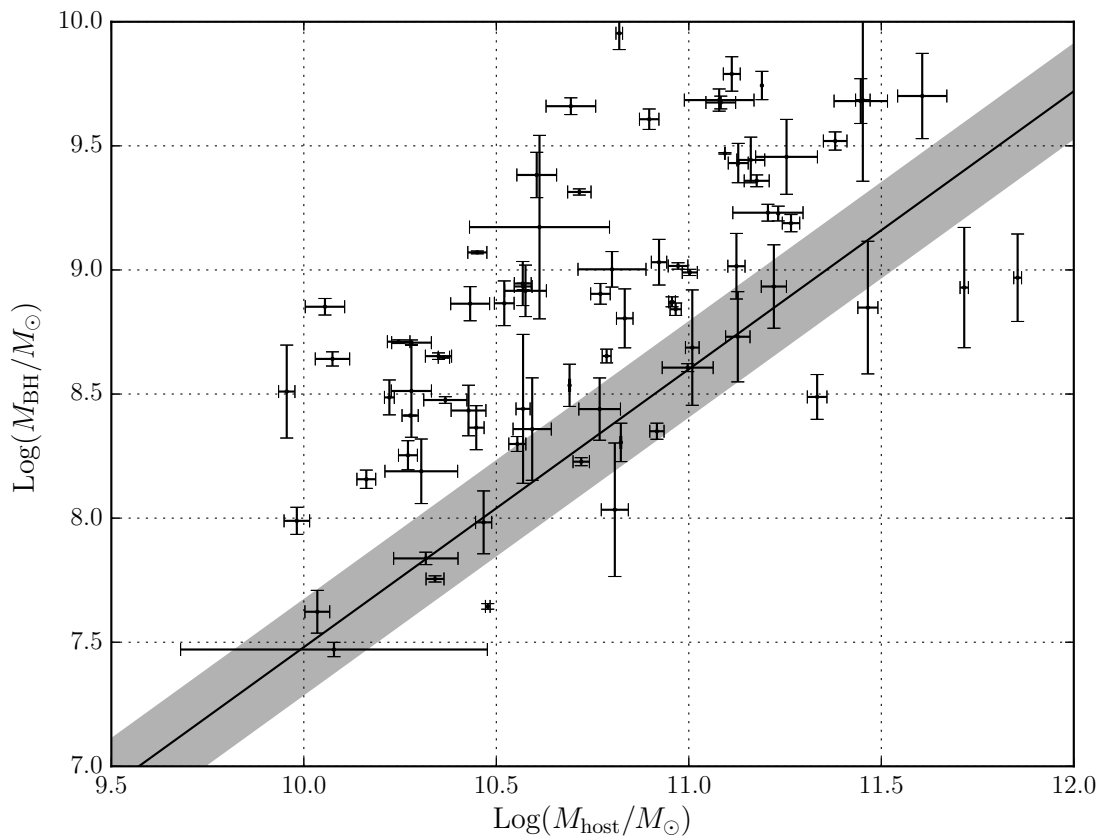}
\caption{The $M_{\rm BH}$ vs.~$M_{\ast}$ evolution. The x-axis is the host galaxy stellar mass. The y-axis is the central black hole mass. The black solid line is the $M_{\rm BH} - M_{\ast}$ scaling relation at $z=0$ from \cite{Bennert2011}. The gray area indicates the $3\,\sigma$ boundary. 42 out of 93 AGNs BH masses are above the $3\,\sigma$ boundary, indicates an evolutionary trend. \label{Fig:mhost-mbh}} 
\end{figure}

\begin{figure}
\plotone{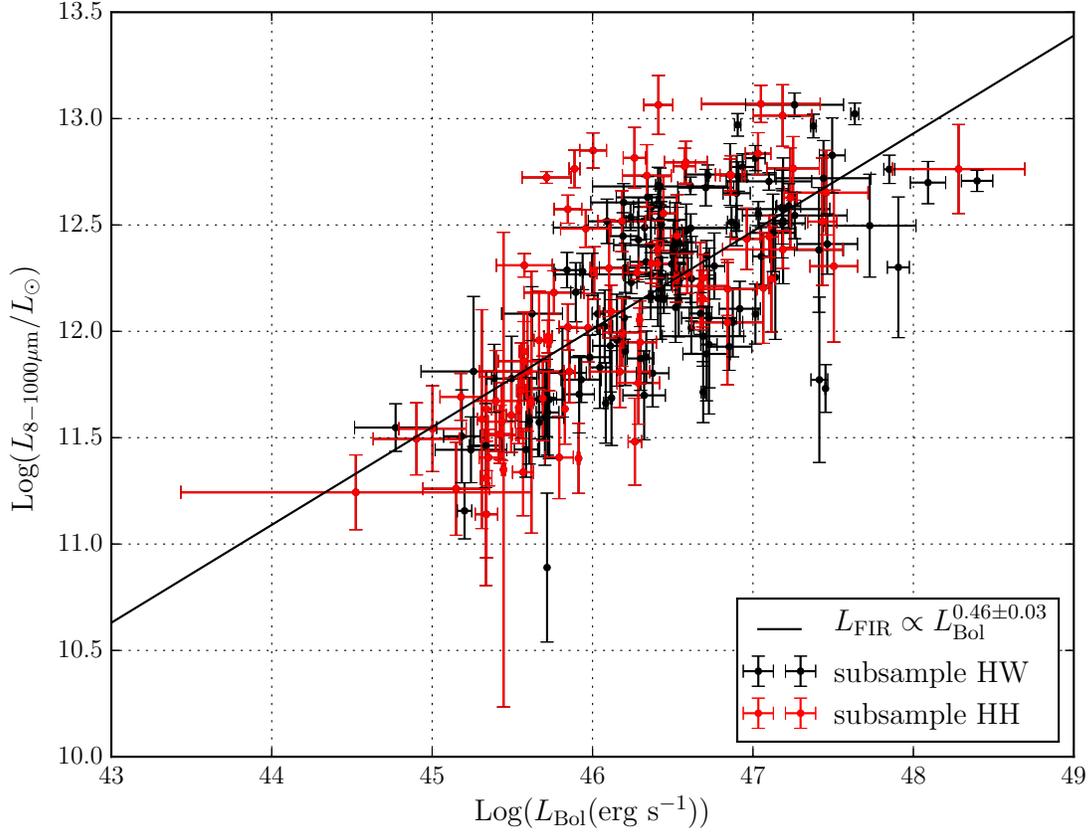}
\caption{The AGN bolometric luminosity vs.~far infrared luminosity of the host. quasars in the subsample HH (quasars with noticeable host) are represented in red, while quasars in the subsample HW (quasars without noticeable host) are represented in black. The solid line indicates a positive relation, where $L_{\rm FIR} \propto L_{\rm BOL}^{0.46\pm0.03}$. The quasars within subsample HH and HW follow the same $L_{\rm Bol}$ vs.~$L_{\rm FIR}$ relation, with the same significant level. \label{Fig:bol-fir}}
\end{figure}

\clearpage

\begin{center}
\begin{longtable}{ll}
\caption{The SED fitting results: table description \label{Tab:sample}}\\
\hline\hline
Column & Description \\
\hline
1& Quasar index\\
2 & Quasar designation: $hhmmss.ss+ddmmss.s$ (J2000.0)\\
3 & Right ascension in decimal degrees (J2000.), taken from HerS catalog\\
4 & Declination in decimal degrees (J2000.0), taken from HerS catalog\\
5 & Redshift, taken from the SDSS DR7 and DR10\\
6 & Host galaxy fraction at 1$\mu$m, $f_g$\\
7 & Uncertainty in $f_g$\\
8 & The internal reddening $A_{int}$. In units of magnitude\\
9 & Uncertainty in $A_{int}$\\
10 & Host galaxy morphological type\\
11 & FIR cold dust temperature $T_{cold}$, in units of Kelvin\\
12 & Uncertainty in $T_{cold}$\\
13 & AGN power-law index at UV/optical $\alpha$ \\
14 & Uncertainty in $\alpha$\\
15 & FIR luminosity integrated from 8$\mu$m to 1000 $\mu$m Log($L_{\rm FIR}$(erg s$^{-1}$))\\
16 & Uncertainty in Log($L_{\rm FIR}$(erg s$^{-1}$))\\
17 & Bolometric luminosity Log($L_{\rm Bol}$(erg s$^{-1}$))\\
18 & Uncertainty in Log($L_{\rm Bol}$(erg s$^{-1}$))\\
19 & SFR Log(SFR($M_{\odot}$/yr))\\
20 & Uncertainty in Log(SFR($M_{\odot}$/yr))\\
21 & Monochromatic luminosity at 1300 \AA, Log($L_{1300}$(erg s$^{-1}$))\\
22 & Uncertainty in Log($L_{1300}$(erg s$^{-1}$))\\
23 & Monochromatic luminosity at 3000 \AA, Log($L_{3000}$(erg s$^{-1}$))\\
24 & Uncertainty in Log($L_{3000}$(erg s$^{-1}$))\\
25 & Monochromatic luminosity at 5100 \AA, Log($L_{5100}$(erg s$^{-1}$))\\
26 & Uncertainty in Log($L_{5100}$(erg s$^{-1}$))\\
27 & Monochromatic luminosity at 250 $\mu$m, Log($L_{250\mu m}$(erg s$^{-1}$))\\
28 & Uncertainty in Log($L_{250\mu m}$(erg s$^{-1}$))\\
29 & Black hole mass, Log($M_{\rm BH}/M_{\odot}$)\\
30 & Uncertainty in Log($M_{\rm BH}/M_{\odot}$)\\
31 & Gas mass, Log($M_{\rm gas}/M_{\odot}$)\\
32 & Uncertainty in Log($M_{\rm gas}/M_{\odot}$)\\
33 & Host galaxy mass, Log($M_{\ast}/M_{\odot}$)\\
34 & Uncertainty in Log($M_{\ast}/M_{\odot}$)\\
35 &  SDSS quasar catalog\\
\hline\hline
\end{longtable}
\end{center}
\clearpage

\begin{sidewaystable}
\begin{center}
\caption{Detailed information for quasars in Figure \ref{Fig:example}. \label{Tab:example}}
\begin{tabular}{cccccccccccc}
\hline\hline
num & names & redshift & $A_{int}$  & $f_{g}$  & hostType & $\alpha$ & $T_{cold}$ \\
\hline
a & 022031.18-010458.2 & 1.64 & $0.1 \pm 0.05$ & $0.0 \pm 0.1$ & none & $-0.09 \pm 0.04$ & $27.9\pm 3.00$ \\
b & 021734.63-002641.9 & 1.56 & $0.0 \pm 0.0$   & $0.0 \pm 0.0$ & none & $0.05\pm 0.01$ & $31.8\pm 3.10$ \\
c & 021857.19-004158.4 & 0.886 & $0.2 \pm 0.05$ & $0.4 \pm 0.1$ & E & $-0.2 \pm 0.4 $& $22.9\pm 2.42$ \\
d & 020837.95-003422.2 & 2.26 & $0.0 \pm 0.04$ & $0.2 \pm 0.1$ & E & $0.08\pm 0.1$ & $41.7 \pm 6.02$ \\
e & 014648.36-002422.4 & 0.804 & $0.1 \pm 0.02$ & $0.1 \pm 0.08$ & S & $0.04 \pm 0.1$ & $22.3 \pm2.30 $\\
f & 021100.99-004401.9 & 1.36 &$ 0.0 \pm 0.0$  & $0.3 \pm 0.05$  & S & $0.3 \pm 0.5$ & $24.3\pm 2.28$ \\
\hline\hline
\end{tabular}
\end{center}
\textbf{Notes.}
\begin{itemize}
\begin{small}
\item[] $A_{int}$, internal extinction in units of magnitude, deduced from the \textquotedblleft quasar-galaxy mixing diagram\textquotedblright. 
\item[] $f_g$, the host galaxy fraction at 1 $\mu$m, also deduced from the \textquotedblleft quasar-galaxy mixing diagram\textquotedblright. 
\item[] $\alpha$, the AGN power-law index at UV/optical
\item[] $T_{cold}$, the FIR cold dust temperature. 
\end{small}
\end{itemize}
\end{sidewaystable}

\clearpage

\begin{table}
\begin{center}
\caption{Sample Summary. \label{Tab:sum}}
\begin{tabular}{c|ccc}
\hline\hline
Sample & Entire sample & Subsample-HH & Subsample-HW\\ 
\hline
redshift & 1.6$\pm$0.77 & $1.4\pm0.83$ & $1.7\pm0.70$\\
LOG($L_{\rm Bol}$ (erg s$^{-1}$))& $46.4\pm0.672$ & $46.2\pm0.709$ & $46.6\pm0.587$\\
LOG($L_{\rm IR}/L_{\odot}$)& $12\pm0.44$ & $12\pm0.50$ & $12\pm0.38$\\
LOG($M_{\rm BH}/M_{\odot}$)& $9.0\pm0.60$ & $8.9\pm0.67$ & $9.1\pm0.53$\\
LOG($M_{\ast}/M_{\odot}$)& N/A & $11\pm0.49$ & N/A\\
LOG($M_{\rm gas}/M_{\odot}$)& $11\pm0.44$ & $11\pm0.48$ & $11\pm0.40$\\
LOG(SFR($M_{\odot}$ Gyr$^{-1}$))& $2.4\pm0.44$ & $2.4\pm0.50$ & $2.5\pm0.38$\\
sSFR (Gyr$^{-1}$)& N/A & $0.53\pm0.48$ & N/A\\
$\tau_{\rm SFR}$ (Gyr) & $0.56\pm0.29$ & $0.64\pm0.37$ & $0.51\pm0.18$\\
LOG$(M_{\rm BH}/M_{\ast}) $ & N/A & $0.022\pm0.031$ & N/A\\
$M_{\rm gas}/(M_{\rm gas} + M_{\ast})$ & N/A & $0.51\pm0.012$ & N/A\\
\hline\hline
\end{tabular}
\end{center}
\end{table}
\clearpage


\end{document}